\documentclass[a4paper]{sab}  

\usepackage{graphicx} 
\usepackage{txfonts,natbib} 
\bibpunct{(}{)}{;}{a}{}{,} 

\usepackage[T1]{fontenc} 
\usepackage{t1enc} 
\usepackage{aeguill} 
\renewcommand\ackname{Acknowledgements.} 
\renewcommand\keywordname{Keywords.} 
\renewcommand\figureptname{Figure}
\renewcommand\tableptname{Table}

\renewcommand\noteaddname{Note added to proofs.}
\idline{Boletim da Sociedade Astron\^omica Brasileira, {\bf 30}, 222-225}{1}
\begin{document} 
\title{Gamma-ray Emission Properties of Four Bright \emph{Fermi}-LAT
AGNs: Implications on Emission Processes} 

\author{Pankaj Kushwaha \inst{1}
        \and 
        Atreyee Sinha \inst{2}
        \and
        Ranjeev Misra \inst{2}
        \and
        K. P. Singh \inst{3}
        \and
        E. M. de Gouveia Dal Pino \inst{1}
        }

\offprints{Pankaj Kushwaha}

\institute{Department of Astronomy (IAG-USP), University of Sao Paulo, Sao Paulo 05508-090, Brazil \\
           \email{pankaj.kushwaha@iag.usp.br}
           \and
           Inter-University Center for Astronomy and Astrophysics, Pune 411007, India
           \and
           Department of Astronomy \& Astrophysics, Tata Institute of Fundamental Research, Mumbai 400005, India
           }
\date{Received 2017 December 8; Accepted 2017 December 8} 

\Abstract 
{The X-ray/Ultraviolet/Optical emission from radio-quiet AGNs, black hole binaries,
and other compact sources, in general, follow a lognormal flux distribution, a linear
rms-flux relation, and a (broken) power-law power spectral densities (PSDs). These
characteristics are normally attributed to the multiplicative combination of fluctuations
in the accretion disk. Similar features have been inferred for some well-observed
blazars in different energy bands, but a systematic study over a long duration is
still missing. Using a continuous gamma-ray light curves over 3-days cadence from
August 2008 – October 2015, we present the first systematic study of these features
in four sources: the FR I radio galaxy NGC 1275 and three blazars- Mrk 421, B2 1520+31
and PKS 1510-089. For all, except Mrk 421, the flux spans $\gtrsim$ 2 orders of
magnitude. For blazars, a log-normal profile describes the flux histograms better
compared to a Gaussian, while none is favored for NGC 1275, the only non blazar source,
suggesting a complex distribution. Regardless of flux histogram profile, the
rms-flux relation is linear for all with PSDs being consistent with a power-law
shot noise spectrum despite hints of breaks. The inferred results are consistent
with the properties of unresolved magnetic reconnection sites, as inferred in the
X-ray emission from the whole Solar disk and the statistical characteristics of
magnetic reconnection based minijets-in-a-jet model. The results, thus, suggest
a strong jet-accretion-disk coupling with energy input from the central source
being distributed over a wide range in time and energy by the reconnection process
depending on the geometry and local physical conditions.}
{restricted due to abstract limit}



\maketitle 

\section{Introduction}
Active Galactic Nuclei (AGNs) represent the class of astrophysical galaxies powered
by accretion onto a super massive black hole (SMBH). Historically divided empirically
based on observational properties, primarily on features in the radio and optical
bands, they exhibit a huge and wide range of characteristics in different energy
bands spectrally, temporally, and spatially \citep{2008NewAR..52..227T}. They emit
high and rapidly variable emission across the entire observable electromagnetic
spectrum and constitute the largest fraction of sources in any extra-galactic surveys.
At $\gamma$-ray energies covered by the Fermi-LAT (Large Area Telescope), they
constitute $\gtrsim$50\% of the total population in the latest LAT source catalog
\citep[3FGL;][]{2015ApJS..218...23A} and $>$ 75\% above 10 GeV \citep[3FHL;]
[]{2017ApJS..232...18A}.

\begin{figure*}[ht]
 \centering
 \resizebox{\hsize}{!}{\includegraphics{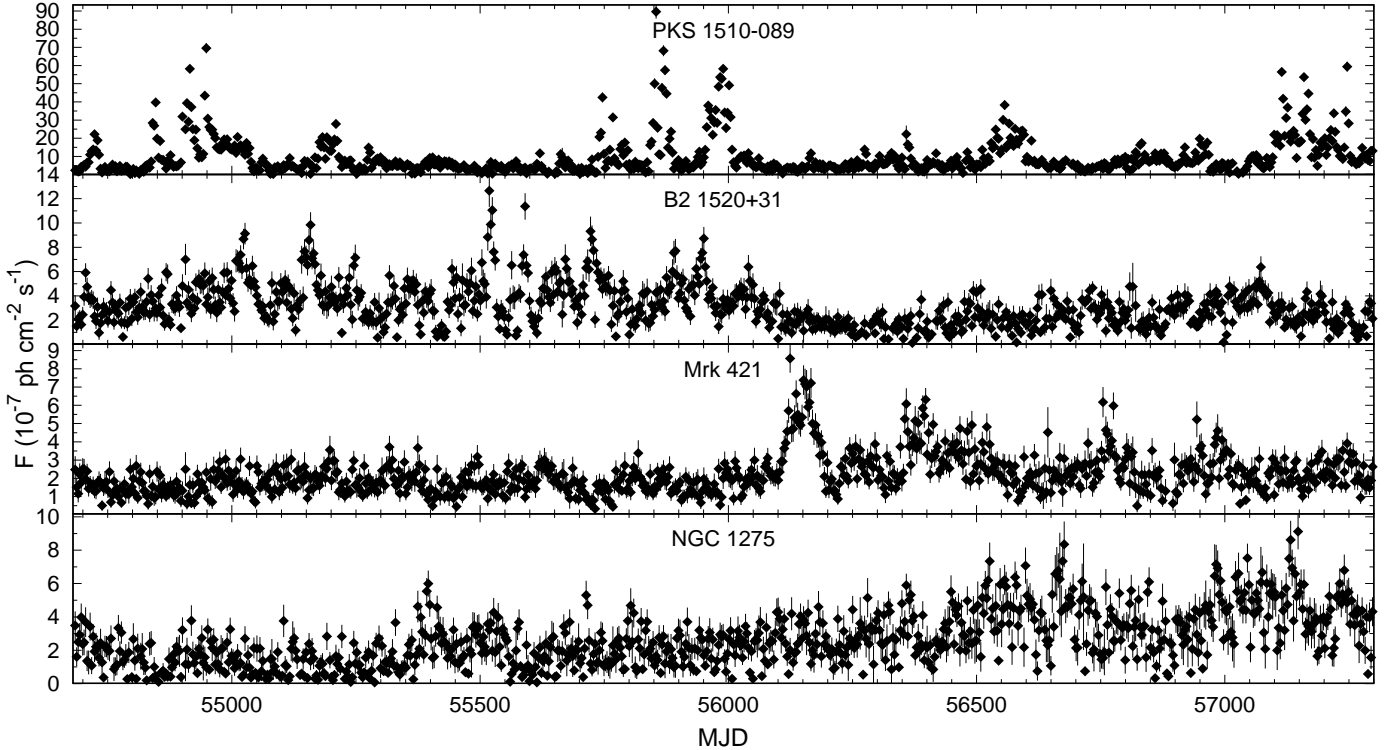}}
 \caption{3-days binned, 0.1-300 GeV \emph{Fermi}-LAT $\gamma$-ray light curve of
 radio-galaxy NGC 1275 and three blazars: PKS 1510-089, B2 1520+32 and Mrk 421 from
 August 5, 2008 to October 5, 2015.}
 \label{fig:lc}
\end{figure*}
One of the defining characteristics of accretion-powered sources is the high, rapid,
and energy-dependent brightness variability \citep{2015SciA....1E0686S,2005MNRAS.359..345U,
2003MNRAS.345.1271V}. Extensive investigation of X-rays emission from compact sources
such as luminous AGNs (most being radio-quiet), galactic black hole binaries (GXBs)
and micro quasars show remarkable phenomenology in statistical properties of emission,
characterized by a log-normal
flux distribution, a linear scaling of intrinsic variability with flux (rms-flux),
and a (broken) power-law power spectral density \citep[PSD;][and references therein]
{2005MNRAS.359..345U,2003MNRAS.345.1271V}. Similar characteristics have been found
in (optical/Ultraviolet/X-ray) emission from other non-compact accreting sources
like young stellar objects, white dwarfs, and cataclysmic variables \citep[and references
therein]{2015SciA....1E0686S}. These remarkable similarities of emission features
across the mass scale have been used to claim that the physics of accretion is
universal and is independent of the accretor physical attributes. The most accepted
explanation of such scale-free features is the multiplicative combination of
fluctuations in the accretion disk \citep{2005MNRAS.359..345U,1997MNRAS.292..679L}.

Presence of magnetic field, even sub-dominant, can substantially modify the
characteristics of a system by introducing multiple scales of coupling and directions.
In fact, the X-ray emission from the whole Solar disk exhibits all the above mentioned
statistical features \citep{2007HiA....14...41Z} but the variability here is the result
of magnetic reconnection process \citep{2017ApJ...836...17A}. This makes the
uniqueness claim ambiguous in the magnetized accretion-powered sources like blazars
and radio galaxies, the sources considered in this study \citep[e.g.][magnetic reconnection scenario for
non-blazars sources]{2015ApJ...802..113K,2005A&A...441..845D}. In these sources, the dominant emission
is non-thermal and is believed be produced within the relativistic jets. But, being
accretion-powered and magnetized, study of these systems offer a potential way to
explore the accretion-disk-jet connection and relative roles of particles and magnetic
field. In addition, a comparative study of physical process in other astrophysical
sources which exhibit similar statistical features may further provide a broader
insights into the jet processes. Here, we present some of the main results and 
summary of our work published in \cite{2017ApJ...849..138K}.

\section{Data}

We have used 7 years of LAT $\gamma$-ray data from August 5, 2008, to October 5, 2015, of four
bright AGNs: radio-galaxy NGC 1275, and three blazars Mrk 421, B2 1520+31 and PKS
1510-089 to investigate the statistical features at $\gamma$-rays. These are the
only AGNs which have a near continuous detection in the LAT on a shortest possible,
uniformly binned timescale of 3-days. The choice of bin-duration is a compromise
between maximizing the data length for PSDs and rms-flux estimation while, at the
same time, avoiding significant non-detections which can bias the histograms. The
extracted light curve is shown in Figure \ref{fig:lc} and the details of work and
data reduction method are published in the \citet{2017ApJ...849..138K}.

\section{Analysis and Results}\label{sec:analysis}

\begin{figure}
 \centering
 \resizebox{\hsize}{!}{\includegraphics{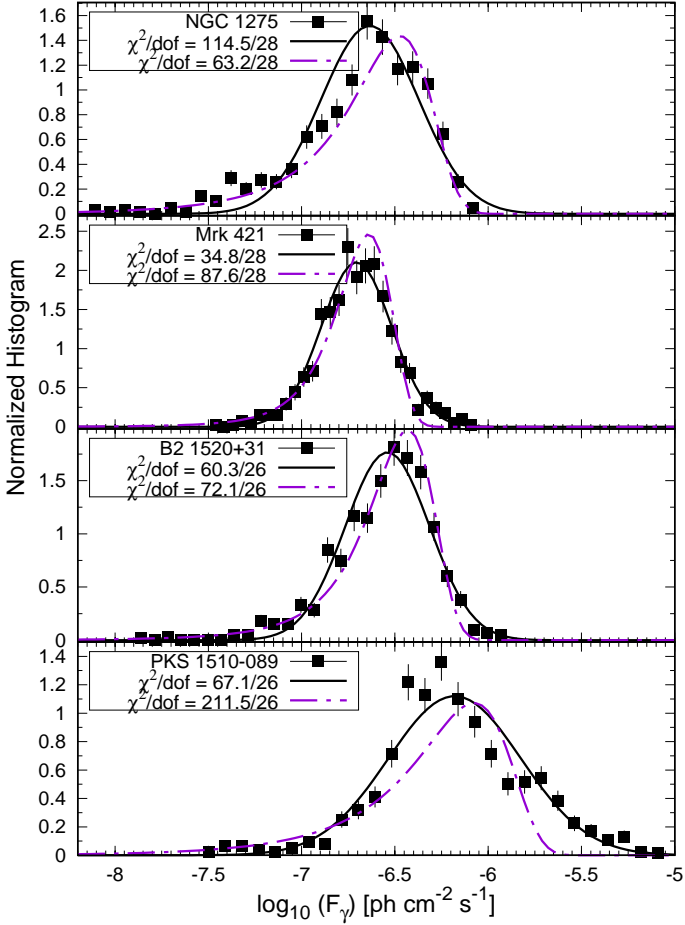}}
 \caption{Flux histogram estimated using the Knuth method. The solid and the dashed-
 dotted  curves are, respectively, the best fit log-normal and Gaussian to the
 histograms (dof = degree of freedom, see \S\ref{sec:analysis})}
 \label{fig:hist}
\end{figure}
We have investigated the statistical properties: flux distribution, PSDs and the rms-flux
relation which are used to study the accretion-powered sources, for the first
time at $\gamma$-ray energies. For highly variable source, a histogram is a unique
tool to search for the presence of scales in the data domain while PSD in the temporal
domain. Figure \ref{fig:hist} shows the $\gamma$-ray flux histograms generated following
Knuth method \citep{2006physics...5197K}. We have fitted the histograms with a log-normal
and a Gaussian profile and the fit statistics is reported in the Figure label, favoring
a log-normal for blazars and a Gaussian for NGC 1275.

\begin{figure}
 \centering
 \resizebox{\hsize}{!}{\includegraphics{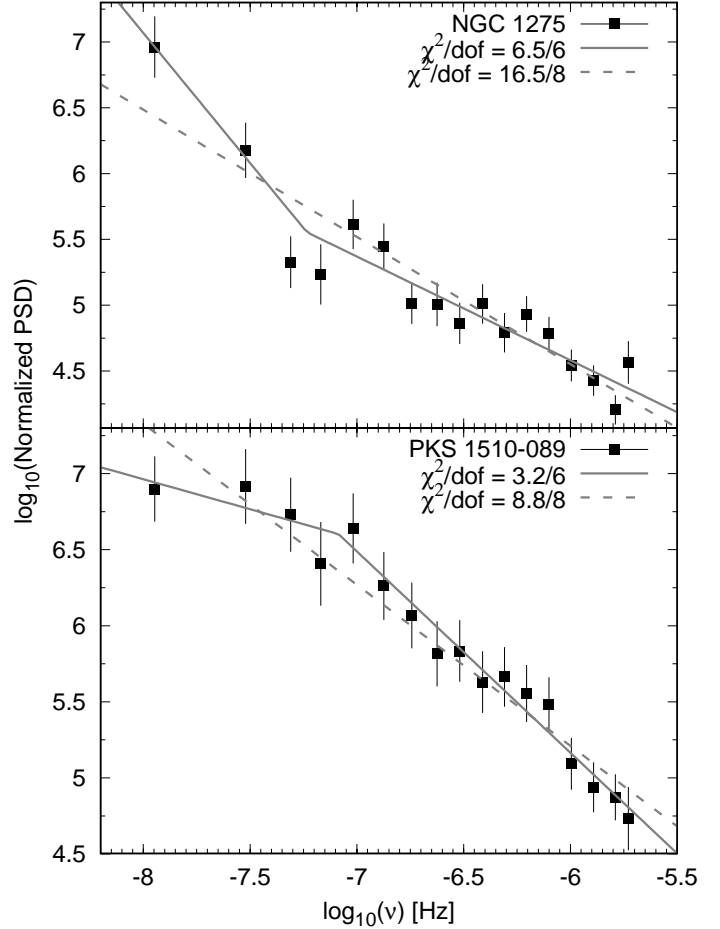}}
 \caption{Gamma-ray PSD of radio-galaxy NGC 1275 and blazar PKS 1510-089. The dashed
 and solid curves are, respectively, the best fit power-law and broken power-law fit
 to the data (see \S\ref{sec:analysis}).}
 \label{fig:psd}
\end{figure}

Figure \ref{fig:psd} shows the PSD of the two sources and the best fit power-law and
broken power-law curve to the data with the fit statistics in the respective labels.
The PSDs are estimated following the method of \citet{2005astro.ph..1215G} while the
error bars are estimated via simulating 1000 light curves using the method of
\citet{1995A&A...300..707T}. The fit favors a broken power-law over power-law interpretation
with similar inferences for the other two sources: Mrk 421 and B2 1520+31 (but see
\S4)

\begin{figure}
 \centering
 \resizebox{\hsize}{!}{\includegraphics{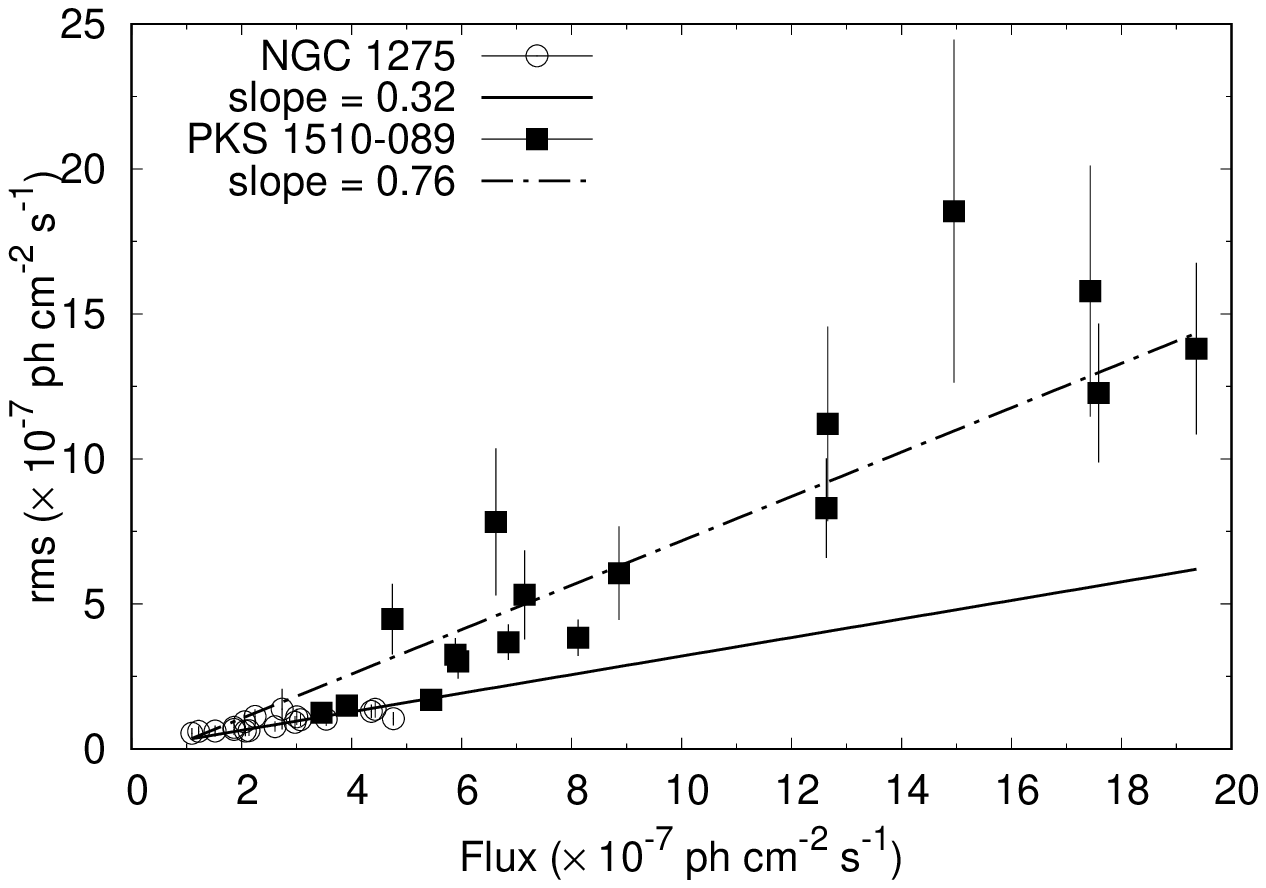}}
 \caption{Intrinsic variability (RMS) as a function of source flux. The dashed-dotted
 solid line represents the best linear fit to the radio-galaxy NGC 1275 and blazar
 PKS 1510-089 with best file slope in the label (see \S\ref{sec:analysis}).}
 \label{fig:rms}
\end{figure}
In figure \ref{fig:rms}, we present the shows intrinsic variability as the function
of source flux for NGC 1275 and PKS 1510-089, estimated using a sequence of 50 data per
rms point from the observed light curve. The error on the rms is calculated
via simulating 1000 light curves as done for the PSD followed by their sampling as per the
observed light curve and then estimating the variance \citep{2003MNRAS.345.1271V}.
The best fitted linear curves with slope values are given in the figure labels. 
Similar linear relation holds for the other two sources with slope consistent with
the NGC 1275 value.

\section{Discussion}
Variability, spectrally and temporally, is one of the few potential tools available 
to study rapidly variable sources which are beyond the resolution limit of any
of the modern observing facilities. Among highly variable astrophysical sources,
accretion-powered sources are the most prominent and have been explored extensively
in different energy bands, especially at X-rays energies \citep{2015SciA....1E0686S,
2005MNRAS.359..345U,2005astro.ph..1215G,2003MNRAS.345.1271V}. The temporal investigations
have found some hallmark statistical features exhibited by these sources: a log-normal flux
distribution with a linear relation with intrinsic variation (rms) and a shot nose
PSD. These features have been claimed to be an imprint of a multiplicative combination of
fluctuations in the accretion-disk. Here, we investigate these features, for the first
time at $\gamma$-ray energies to gain insights into jet-disk connection and jet processes.

The systematic investigation reveals many interesting results with broad implications
on the jet processes. In the flux domain,
all the histogram show a prominent peak with extremes separated by $\gtrsim$
2 order of magnitudes except for Mrk 421. The width of the distribution around
the peak is similar for all though there is a huge tail at the lower energy end
\citep{2017ApJ...849..138K}. The fit results for histograms, mentioned in labels of Fig.
\ref{fig:hist} show that a log-normal profile is favored for blazars while a Gaussian
is favored for the sole non-blazar source NGC 1275. However, an in-depth statistical
analysis by \citet{2017ApJ...849..138K} has shown that neither of the distributions
adequately describes the NGC 1275 histogram. Surprisingly, the rms-flux relation
is linear for all and so is the variance with respect to source flux state except for
PKS 1510-089 (see Fig. \ref{fig:rms}). The linearity of rms-flux relation is contrary
to the inferences from other accretion powered sources e.g. radio-quiet AGNs, BXBs,
YSOs, CV etc. where lognormality is argued for the linear rms-flux relation and is
attributed to the multiplicative combination of the fluctuations in the accretion
disk. In the Fourier domain, the PSDs exhibit a shot noise spectra (see Fig. \ref{fig:psd}),
typical of accretion powered sources. Though a broken power-law fit seems to provide a
better description of the PSD, the quality of data suggests breaks at many frequencies
for the single broken power-law, while we have reported only the one which resulted
in lowest $\chi^2$ statistics. Thus, the nature of the best fit suggests that a more
in-depth analysis is needed to statistically claim the presence of breaks within
the considered duration if any. 

The inferred results are broadly consistent with the statistical properties of
the magnetic reconnection powered minijets-in-a-jet model \citep{2009MNRAS.395L..29G,
2012A&A...548A.123B,2012MNRAS.426.1374C}. In this model, emission from identical
but randomly oriented emission regions follows Pareto distribution
\citep{2005PApGe.162.1187Z}. It predicts a huge range of histogram profiles, from
a power-law in case of a single emission region to one that can be interpreted as
the log-normal in case of contribution from large number of emission regions with
a linear rms-flux relation for all \citep{2012A&A...548A.123B}, as inferred for
the $\gamma$-ray emission here (skewed or log-normal). The inferences are also
consistent with the statistics of some of the physical quantities associated with
the X-ray emission from the whole Solar disk and coronal mass ejections (CMEs)
\citep{2007HiA....14...41Z}, both of which are known to be the result of magnetic
reconnection. The results, combined with the highly magnetized nature of these
sources suggest that magnetic reconnection is the dominant process responsible
for powering the jet emission. Within this, the skewed distribution of NGC 1275
suggests that fewer regions are contributing to the emission in comparison with
blazars. Since the central compact source is the main source of power, the inferences
suggest that the magnetic reconnection may be a result of fluctuations in the
accretion disk and/or it could be intrinsic to the jet as a result of
non-linear dissipative processes within it. Another possibility could be that it is an
imprint of magnetic reconnection in the corona, the likely location of the jet base,
as argued in \citet{2005A&A...441..845D} for sources having non-boosted emission
\citep{2015ApJ...802..113K}. The latter scenario is interesting in the sense that
if these imprints flow down the relativistic jet, the overall radiative output will
be boosted and the boosting will also makes it appear as if more regions are
contributing to the emission, thereby shifting the histogram profile closer 
to the log-normal, as is the case with blazars. It should be noted that a boosting
factor of $\sim$ 10 can produce flux similar to the blazars.

\section{Summary}
Investigation of characteristics like flux distribution, rms-flux relation, and PSD
at $\gamma$-ray energies which are used to characterize accretion-powered source 
show broadly similar yet interesting results. We find that the flux distribution exhibit complex shape with
log-normal being favored over Gaussian for blazars but none for NGC 1275. Irrespective
of the flux distribution, the rms-flux relation is linear for all and the PSDs are
typical of accretion-powered sources i.e. shot noise profile. The linearity of rms-flux
relation irrespective of the histogram profile disfavors a strong accretion disk
contribution as generally claimed for accretion-powered sources, though there
may be contributions. Instead, the features are consistent with the statistical
properties of magnetic reconnection powered
minijets-in-a-jet model and some characteristics of the quantities associated with
the Sun like X-ray emission from whole Solar disk and CMEs, thereby favoring magnetic
reconnection as the source powering relativistic jets. This is also consistent with
the general emission characteristics of these sources such as a non-thermal broadband
spectrum and the rapid variability. The analysis suggests that magnetic reconnection
may be an imprint of accretion disk fluctuations on the jet or effect of dynamics
within the jet and/or could be an imprint of the corona. 

\begin{acknowledgements} 
PK is supported by grants from the Brazilian agency FAPESP 2015/13933-0. EMGDP also
acknowledges support from  FAPESP (2013/10559-5) and CNPq (306598/2009-4) grants.
\end{acknowledgements}

\end{document}